\documentclass[aip,jcp,preprint,groupedaddress, showpacs]{revtex4-1}
 
\usepackage{graphicx}
\usepackage{dcolumn}
\usepackage{bm}
\usepackage{amsmath}
\usepackage{textcomp}
\usepackage{longtable}
\usepackage{url}
\usepackage{tabularx}
\usepackage{threeparttable}
\usepackage{multirow}
\usepackage{mathtools}

\begin{document}

\title{The crystalline structure of orthorhombic SrRuO$_3$: Application of hybrid scheme to the density functionals revised for solids}

\author{\v{S}. Masys }
\email[Corresponding author. Email address: ]{Sarunas.Masys@tfai.vu.lt}

\author{V. Jonauskas}

\affiliation{Institute of Theoretical Physics and Astronomy, Vilnius University, Saul\.{e}tekio Avenue 3, LT-10257 Vilnius, Lithuania}

\date{\today}

\begin{abstract}

The crystalline structure of ground-state orthorhombic SrRuO$_3$ is reproduced by applying hybrid density functional theory scheme to the functionals based on the revised generalized-gradient approximations for solid-state calculations. The amount of Hartree-Fock (HF) exchange energy is varied in the range of $5-20\%$ in order to systematically ascertain the optimum value of HF mixing which in turn ensures the best correspondence to the experimental measurements. Such investigation allows to expand the set of tools that could be used for the efficient theoretical modelling of, for example, only recently stabilized phases of SrRuO$_3$.          
 
\end{abstract}



\maketitle

\section{Introduction}

Strontium ruthenate SrRuO$_3$ is a perovskite-structured conductive ferromagnet which upon heating undergoes a series of phase transformations: orthorhombic ($Pbnm$) $\xrightarrow{820 \text{ } \text{K}}$ tetragonal ($I4/mcm$) $\xrightarrow{950 \text{ } \text{K}}$ cubic ($Pm\bar{3}m$) [\onlinecite{kennedy_76}]. Nowadays SrRuO$_3$ fascinates researchers because of its pivotal role of being a key integrant for fabrication of oxide heterostructures and superlattices, which in turn have the potential to contribute to new functionalities in electronics and spintronics [\onlinecite{koster_68}]. However, by looking back from a 50-years perspective one can find some 1000 papers spanning the physics, materials science, and applications of SrRuO$_3$ in its bulk and thin-film form, and notice the fact that interest in this material continuously increases. 

A recent observation that pairwise differences between the results of modern solid-state codes, based on density functional theory (DFT) approaches, are comparable to those between different high-precision experiments [\onlinecite{lejaeghere_115}] sheds a new light on the predictive potential of the first-principles simulations. In our previous paper [\onlinecite{masys_116}], we have carefully benchmarked a bunch of DFT functionals -- including local density approximation, generalized-gradient approximations (GGAs), and hybrids -- in order to identify the ones that are the best at reproducing the crystalline structure of ground-state orthorhombic SrRuO$_3$. The importance of such calculations has recently grown to a new level due to the experimental breakthrough in stabilizing tetragonal and monoclinic phases of SrRuO$_3$ at room temperature -- the lack of the precision while determining the exact space group symmetry for these stabilized systems paves the way to exploit the predictive power of DFT simulations. But in order to take advantage of it, firstly one has to be aware of the functionals that could potentially lead to precise reproduction of various crystalline structures of SrRuO$_3$. Our observations, based on the direct comparison to the low-temperature experimental data of orthorhombic symmetry, indicate that a hybrid scheme combined with the GGAs revised for solids is the most appropriate tool for the accurate description of the external (lattice constants and volume) and internal (tilting and rotation angles together with internal angles and bond distances of RuO$_6$ octahedra) structural parameters simultaneously. However, amount of Hartree-Fock (HF) exchange energy smaller than the standard $25\%$ should be preferred, most likely $16\%$ as in B1WC [\onlinecite{b1wc_51}] or so. That said, we find it important to extend our previous study by investigating the influence of HF exchange in the range of $5-20\%$ and thus in a {\it systematic} fashion determine the best option for SrRuO$_3$. What is more, our goal is to include all three revised GGAs for solids, namely, PBEsol [\onlinecite{pbesol_18}], SOGGA [\onlinecite{sogga_19}], and WC [\onlinecite{wc_20}], instead of focusing on a single one of them. The provided recommendations could be useful for the future research of SrRuO$_3$ by expanding the suitable set of tools for the efficient theoretical modelling.   
    
\section{Computational details}

In this work, the ferromagnetic state of low-temperature orthorhombic ($Pbnm$) SrRuO$_{3}$ was simulated using CRYSTAL14 code [\onlinecite{crystal14_54}] which employs a linear combination of atom-centered Gaussian orbitals. The small-core Hay-Wadt pseudopotentials [\onlinecite{hay_59}] were utilized to describe the inner-shell electrons ($1s^{2}2s^{2}2p^{6}3s^{2}3p^{6}3d^{10}$) of Sr and Ru atoms. The valence part of the basis set for Sr ($4s^{2}4p^{6}5s^{2}$) was taken from the SrTiO$_{3}$ study [\onlinecite{piskunov_60}], while the valence functions for Ru ($4s^{2}4p^{6}4d^{7}5s^{1}$) were adopted from our previous work on non-stoichiometric SrRuO$_{3}$ [\onlinecite{masys_61}]. Concerning the oxygen atom, all-electron basis set was applied from the calcium carbonate study [\onlinecite{valenzano_62}].     
   
The default values were chosen for most of the technical setup while performing full geometry optimization -- the details can be found in CRYSTAL14 user's manual [\onlinecite{crystal14_55}]. However, in terms of atomic units, a parameter that defines the convergence threshold on total energy and five parameters that define truncation criteria for bielectronic integrals were tightened to ($10^{-8}$) and ($10^{-8}$, $10^{-8}$, $10^{-8}$, $10^{-8}$, and $10^{-16}$), respectively. Truncation was made according to the overlap-like criteria: when the overlap between two atomic orbitals was smaller than $10^{-x}$, the corresponding integral was disregarded or evaluated in a less precise way. The allowed root-mean-square values of energy gradients and nuclear displacements were correspondingly set to ($6 \cdot 10^{-5}$) and ($1.2 \cdot 10^{-4}$). In order to improve the self-consistence field convergence, the Kohn-Sham matrix mixing technique (at $80\%$) together with Anderson's method [\onlinecite{anderson_64}], as proposed by Hamman [\onlinecite{hamann_65}], were applied. The reciprocal-space integration was performed with the shrinking factor of 8 that resulted in 125 independent \textbf{\textit{k}} points in the first irreducible Brillouin zone.

Within the employed hybrid scheme, the exchange-correlation energy may be given in the form
\begin{equation}
E_{\text{XC}}^{\text{Hybrid}}=aE_{\text{X}}^{\text{HF}}+(1-a)E_{\text{X}}^{\text{GGA}}+E_{\text{C}}^{\text{PBE}},
\label{eq:equ1}
\end{equation}
where $E_{\text{X}}^{\text{GGA}}$ stands for the exchange energy of PBEsol, SOGGA, or WC approaches, whereas $E_{\text{C}}^{\text{PBE}}$ represents the correlation part of PBE functional [\onlinecite{pbe_57}]. Mixing parameter $a$ that controls the amount of HF exchange energy $E_{\text{X}}^{\text{HF}}$ was varied from 0.05 to 0.2.

\section{Results and discussion}

\begin{figure}
\centering
{
\includegraphics[scale=0.75]{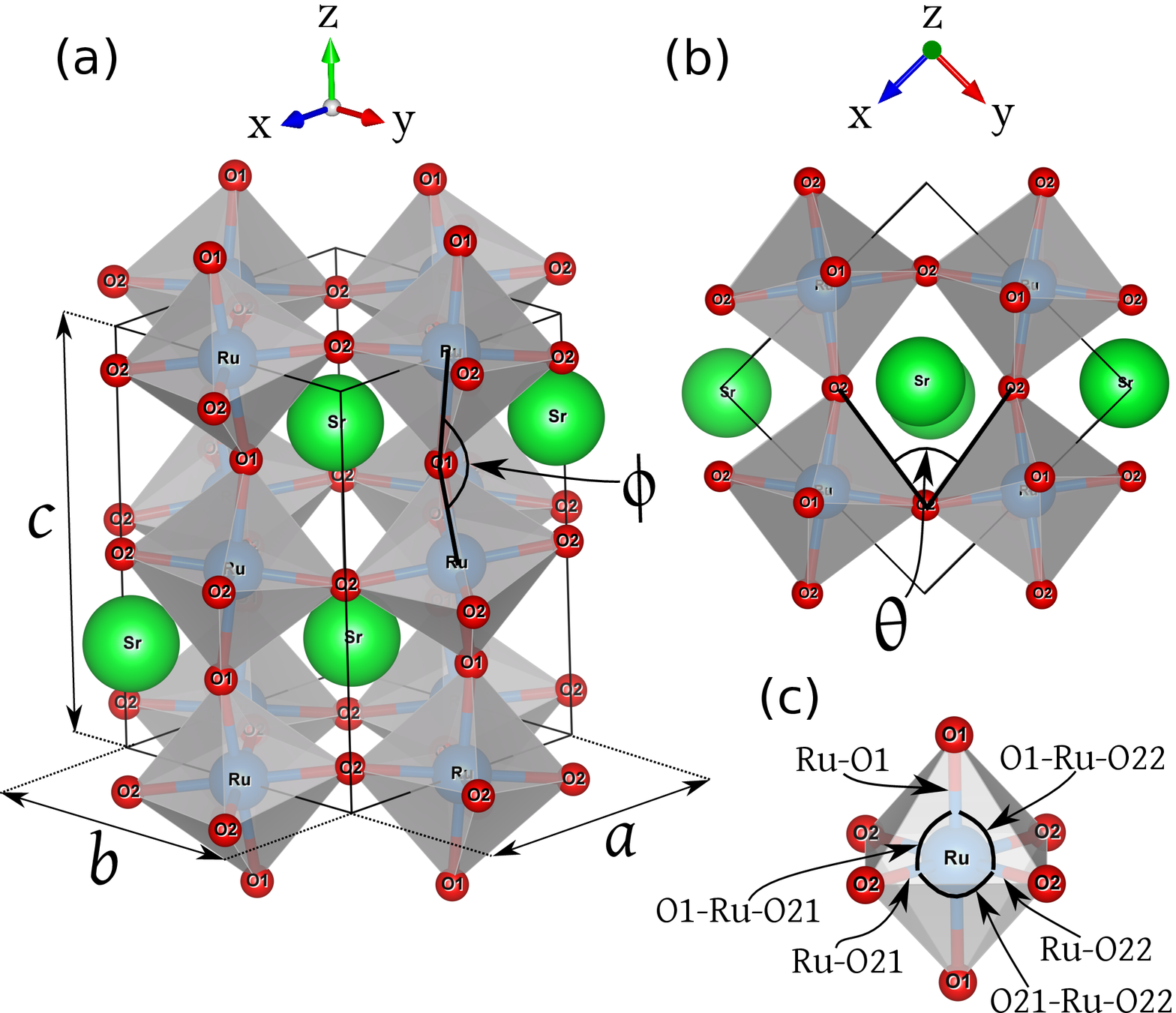}
}
\caption{\label{fig1}
Schematic representation of (a) the crystalline structure of ($Pbnm$) SrRuO$_3$, (b) its top view, and (c) octahedral parameters. Notation O1 and O2 labels oxygen atoms at the apical and planar positions of the RuO$_6$ octahedra, respectively. The drawings were produced with the visualization program VESTA [\onlinecite{vesta_67}].}
\end{figure}

\begin{figure}
\centering
{
\includegraphics[scale=0.58]{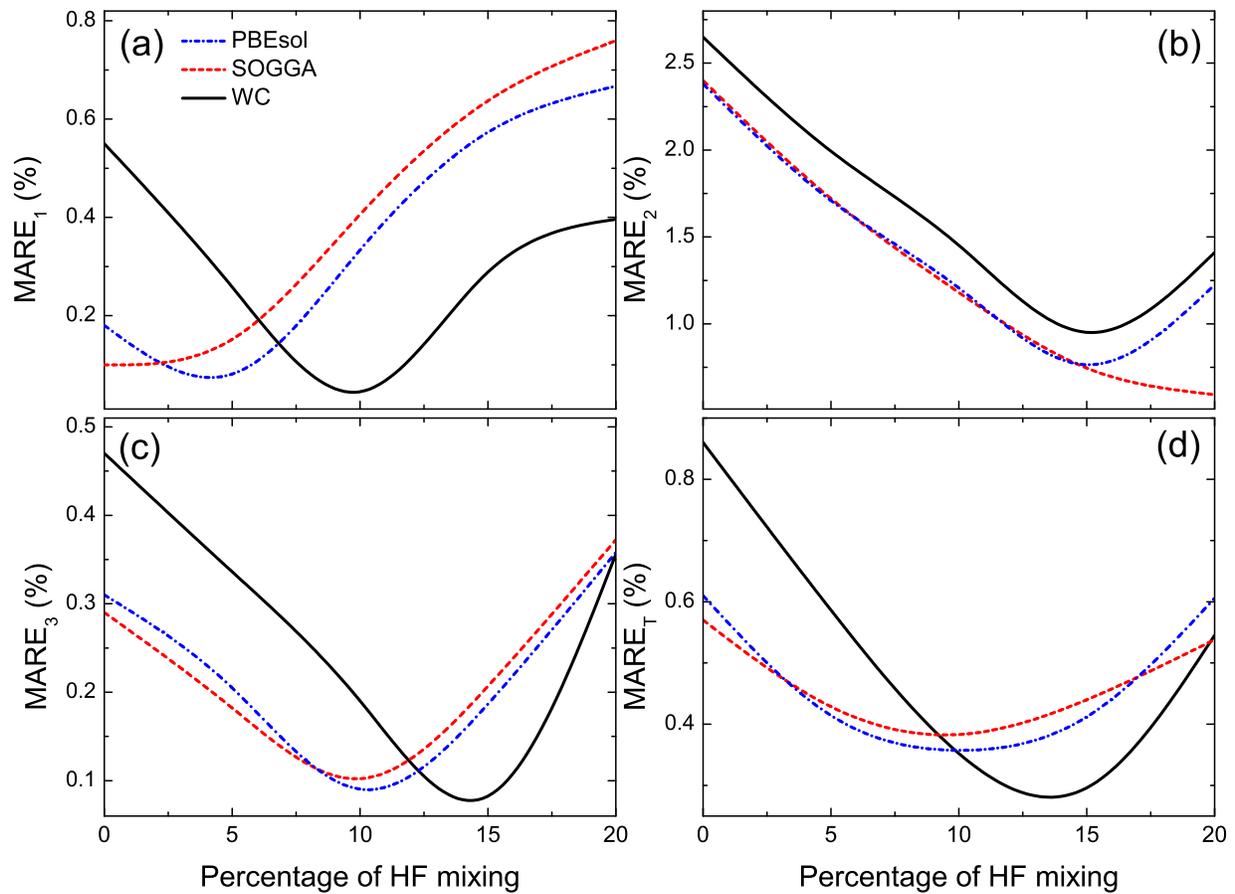}
}
\caption{\label{fig2}
Influence of the amount of HF mixing on (a) MARE$_1$, (b) MARE$_2$, (c) MARE$_3$, and (d) MARE$_{\text{T}}$. The presented curves were smoothed by using a cubic spline interpolation. 
}
\end{figure}

The geometry of ground-state orthorhombic SrRuO$_3$ is depicted in Fig. \ref{fig1}. The equilibrium structural parameters calculated using PBEsol, SOGGA, and WC functionals are given in corresponding Tables \ref{tab1}, \ref{tab2}, and \ref{tab3}. The mean absolute relative errors (MAREs) were evaluated according to the expression
\begin{equation}
\text{MARE} = \frac{100}{n}\displaystyle\sum\limits_{i=1}^{n}{\bigg \vert \frac{p_{i}^{\text{Calc.}}-p_{i}^{\text{Expt.}}}{p_{i}^{\text{Expt.}}} \bigg \vert},
\label{eq:equ14}
\end{equation}
in which $p_{i}^{\text{Calc.}}$ and $p_{i}^{\text{Expt.}}$ are the calculated and experimental values of the considered parameter, respectively. A visual representation of MAREs dependence on the percentage of HF mixing can be found in Fig. \ref{fig2}. For the sake of accuracy, the low-temperature experimental data, also presented in Tables \ref{tab1}-\ref{tab3}, were taken as an arithmetic average of the results obtained from the 1.5 K [\onlinecite{bushmeleva_66}] and 10 K [\onlinecite{lee_66a}] neutron diffraction measurements. It should also be mentioned that no zero-point anharmonic expansion (ZPAE) corrections on the experimental data were applied, since our previous non-magnetic calculations [\onlinecite{masys_104}] indicate that ZPAE correction for the lattice constant of cubic SrRuO$_3$ reaches at most $\sim0.13 \%$ and therefore can be treated as negligible.

An analysis of Fig. \ref{fig2} (a) reveals that the performance of SOGGA functional in reproducing lattice constants and volume is already optimum and addition of HF exchange only worsens the results. However, a small amount ($\sim 5\%$) of $E_{\text{X}}^{\text{HF}}$ appears to be favourable for PBEsol approach which shows a slight improvement in MARE$_1$ reducing it from $0.18 \%$ to $0.08 \%$. A more pronounced amelioration can be noticed for WC functional, since its MARE$_1$ decreases from $0.55 \%$ to $0.05 \%$ at $10\%$ of HF mixing. On the whole, in the range of $a = 0-0.1$ all three revised GGAs are able to yield MARE$_1$ values of $0.1 \%$ or even less, and it can be considered as a truly impressive result. But despite that, a completely different trend is observed in Fig. \ref{fig2} (b) where tilting and rotation angles of RuO$_6$ octahedra are taken into account. One can note that here at least $\sim 15 \%$ of HF mixing is necessary for WC and slightly less for PBEsol and SOGGA in order to get below a so-called satisfactory threshold of MAREs set to $1 \%$ in our previous paper [\onlinecite{masys_116}]. The higher MARE$_2$ values compared to the errors of lattice constants and volume may be explained by the fact that variations in tilting and rotation angles involve very subtle energy changes which are much more harder to deal with. A somewhat different behaviour of SOGGA functional in comparison to PBEsol and WC at $20 \%$ of HF exchange allows to decrease MARE$_2$ to $0.59 \%$ showing that the range of $a = 0.15-0.2$ seems to be the most appropriate choice for the tilting and rotation angles. This observation is perfectly consistent with the result of mB1WC [\onlinecite{masys_116}] -- a combination of WC exchange, PBE correlation, and $16 \%$ of HF mixing -- which is a bit lower ($0.84 \%$) compared to the MARE$_2$ value of WC at $15 \%$ of HF exchange ($0.95 \%$). But amounts of $E_{\text{X}}^{\text{HF}}$ larger than $20 \%$ should not lead to a further improvement though, at least for PBEsol and WC approximations.

Similarly to lattice constants and volume, bond distances and bond angles of RuO$_6$ octahedra are also reproduced with satisfactory accuracy using pure GGA scheme. From Fig. \ref{fig2} (c) and Tables \ref{tab1}-\ref{tab3} one can note that PBEsol, SOGGA, and WC functionals alone achieve MARE$_3$ $< 0.5 \%$, however, additional $10 \%$ of HF mixing for PBEsol and SOGGA and $15 \%$ for WC allow to improve MARE$_3$ values to $0.09 \%$, $0.1 \%$, and $0.08 \%$, respectively. Therefore, it becomes obvious that the range of $a = 0.1-0.15$ is a priority option for the most accurate description of RuO$_6$ geometry. Interestingly, the same tendency also holds for the overall performance of the functionals represented by variation of MARE$_\text{T}$ in Fig. \ref{fig2} (d). Here, MARE$_\text{T}$ value of WC drops from $0.86 \%$ to $0.3 \%$ as the amount of HF exchange is increased up to $15 \%$, while for PBEsol and SOGGA approaches the improvement is not that impressive but still noticeable -- from $0.61 \%$ to $0.36 \%$ for the former and from $0.57 \%$ to $0.38 \%$ for the latter at $10 \%$ of HF mixing. These findings clearly indicate that hybrid scheme has a positive impact on the overall results of all three considered functionals, most likely due to the reduction of self-interaction error which stems from the fact that electrons are allowed to spuriously interact with themselves within GGA framework. An optimum value of HF mixing parameter $a$ falls in the range of $0.1-0.15$, and it is definitely smaller than the typical one of 0.25 usually applied in the first-principles calculations.

\section{Conclusions}

In this study, we have {\it systematically} investigated the influence of $5-20\%$ of HF exchange on the performance of revised GGAs for solids -- PBEsol, SOGGA, and WC -- in reproducing the crystalline structure of ground-state orthorhombic SrRuO$_{3}$. The structural parameters of the system were distinguished into three categories: (a) lattice constants and volume, (b) tilting and rotation angles of RuO$_6$ octahedra, and (c) internal angles and bond distances within RuO$_6$ octahedra. The obtained results indicate that optimum amounts of HF mixing, which ensure the smallest deviations from the experimental measurements, for the corresponding categories fall in the range of (a) $0-10\%$, (b) $15-20\%$, and (c) $10-15\%$. The overall performance of the tested functionals in reproducing structural parameters in all three categories yields deviations smaller than $0.4 \%$, namely, $0.3 \%$ for WC at $15 \%$ of HF exchange and $0.36 \%$ for PBEsol with $0.38 \%$ for SOGGA at $10 \%$ of HF mixing. Thus, in case of the full reproduction of SrRuO$_{3}$ geometry, $10-15\%$ of HF exchange can be considered as the recommended amount for the revised GGA frameworks. These findings expand the available set of tools for theoretical simulations of SrRuO$_{3}$ by revealing that PBEsol and SOGGA approaches can also be combined with HF exchange as efficiently as our previously studied mB1WC scheme based on WC approximation.         

\begin{acknowledgments}
The authors are thankful for the HPC resources provided by the ITOAC of Vilnius University. 
\end{acknowledgments} 

\providecommand{\noopsort}[1]{}\providecommand{\singleletter}[1]{#1}%

\bibliography{paper_masys}

\begin{table}
\caption{\label{tab1}
Structural parameters of orthorhombic ($Pbnm$) SrRuO$_{3}$ calculated within PBEsol framework and compared to the experimental data. Lattice constants $a$, $b$, and $c$ together with bond distances Ru-O1, Ru-O21, and Ru-O22 are given in \AA, volume $V$ is given in \AA$^{3}$, angles $\phi$, $\theta$, O1-Ru-O21, O1-Ru-O22, and O21-Ru-O22 are given in degrees. MARE (in $\%$) stands for the mean absolute relative error: MARE$_1$ is evaluated for $a$, $b$, $c$, and $V$; MARE$_2$ for $\phi$ and $\theta$; MARE$_3$ for Ru-O1, Ru-O21, Ru-O22, O1-Ru-O21, O1-Ru-O22, and O21-Ru-O22; MARE$_\text{T}$ denotes the total MARE of all 12 structural parameters. The numbers in brackets (in $\%$) represent absolute relative errors for each structural parameter.  }
\begingroup                           
\renewcommand{\arraystretch}{0.74}    
\footnotesize
\begin{tabularx}{\textwidth}{>{\hsize=1.0\hsize}X>{\centering\arraybackslash\hsize=1.0\hsize}X>{\centering\arraybackslash\hsize=1.0\hsize}X>{\centering\arraybackslash\hsize=1.0\hsize}X>{\centering\arraybackslash\hsize=1.0\hsize}X
>{\centering\arraybackslash\hsize=1.0\hsize}X>{\centering\arraybackslash\hsize=1.0\hsize}X}
 \hline \hline  
                   & \multirow{2}{*}{PBEsol}     &  \multicolumn{4}{c}{PBEsol with percentage HF mixing}    & \multirow{2}{*}{Expt.}    \\ 
                                                               \cline{3-6}
                   &                             &  5$\%$   & 10$\%$  & 15$\%$   & 20$\%$   &      \\
\hline
$a$                &   5.568                     &  5.563   &   5.555 &    5.545 &   5.534  &  5.566   \\
                   &   (0.05)                    &  (0.04)  & (0.19)  &   (0.38) &  (0.56)  &          \\
$b$                &   5.538                     &  5.523   &   5.513 &  5.504   &  5.518   &   5.531  \\
                   &  (0.13)                     &   (0.12) & (0.32)  &  (0.48)  &   (0.22) &          \\
$c$                &   7.858                     &  7.845   &   7.831 &  7.821   &  7.801   &   7.844  \\
                   &  (0.18)                     &   (0.01) & (0.16)  &  (0.29)  &   (0.55) &          \\
$V$                &  242.30                     &  241.08  &  239.83 &  238.68  &  238.22  &   241.44 \\
                   &  (0.36)                     &  (0.15)  & (0.67)  &  (1.14)  &   (1.33) &          \\
$\phi$             &   159.98                    &  160.31  &  160.85 &  161.51  &  160.35  &   161.97 \\
                   &  (1.23)                     &   (1.02) & (0.69)  &  (0.28)  &   (1.00) &          \\
$\theta$           &   74.43                     &  75.31   &   75.83 &  76.19   &  76.04   &   77.16  \\
                   &  (3.54)                     &   (2.40) & (1.72)  &  (1.25)  &   (1.45) &          \\
Ru-O1              &  1.995                      &  1.991   &   1.986 &  1.981   &  1.979   &   1.986  \\
                   &  (0.46)                     &  (0.25)  & (0.01)  &   (0.23) &  (0.32)  &          \\
Ru-O21             &  1.999                      &  1.993   &   1.988 &  1.982   &  1.992   &   1.986  \\
                   &  (0.63)                     &  (0.33)  & (0.05)  &   (0.22) &  (0.29)  &          \\  
Ru-O22             &  1.997                      &  1.991   &   1.986 &  1.981   &  1.973   &   1.987  \\
                   &  (0.50)                     &  (0.21)  & (0.07)  &   (0.33) &  (0.72)  &          \\
O1-Ru-O21          &  90.20                      & 90.19    &  90.19  &   90.19  & 90.52    &   90.25  \\
                   &  (0.06)                     & (0.07)   & (0.07)  &    (0.07)&  (0.30)  &          \\
O1-Ru-O22          &   90.38                     &  90.45   &   90.46 &  90.45   &  90.05   &   90.31  \\
                   &  (0.07)                     & (0.15)   & (0.16)  &   (0.15) &   (0.30) &          \\
O21-Ru-O22         &   91.21                     &  91.28   &   91.26 &  91.20   &  90.88   &   91.08  \\
                   &  (0.14)                     & (0.22)   & (0.20)  &   (0.12) &   (0.22) &          \\
MARE$_1$           &  0.18                       &   0.08   &   0.33  &   0.57   &   0.67   &          \\
MARE$_2$           &  2.38                       &   1.71   &   1.21  &   0.76   &   1.22   &          \\ 
MARE$_3$           &  0.31                       &   0.20   &   0.09  &   0.19   &   0.36   &          \\ 
MARE$_\text{T}$    &  0.61                       &   0.41   &   0.36  &   0.41   &   0.61   &          \\    

\hline \hline
\end{tabularx}
\endgroup
\end{table}  

\begin{table}
\caption{\label{tab2}
Structural parameters of orthorhombic ($Pbnm$) SrRuO$_{3}$ calculated within SOGGA framework and compared to the experimental data. Lattice constants $a$, $b$, and $c$ together with bond distances Ru-O1, Ru-O21, and Ru-O22 are given in \AA, volume $V$ is given in \AA$^{3}$, angles $\phi$, $\theta$, O1-Ru-O21, O1-Ru-O22, and O21-Ru-O22 are given in degrees. MARE (in $\%$) stands for the mean absolute relative error: MARE$_1$ is evaluated for $a$, $b$, $c$, and $V$; MARE$_2$ for $\phi$ and $\theta$; MARE$_3$ for Ru-O1, Ru-O21, Ru-O22, O1-Ru-O21, O1-Ru-O22, and O21-Ru-O22; MARE$_\text{T}$ denotes the total MARE of all 12 structural parameters. The numbers in brackets (in $\%$) represent absolute relative errors for each structural parameter.  }
\begingroup                           
\renewcommand{\arraystretch}{0.74}    
\footnotesize
\begin{tabularx}{\textwidth}{>{\hsize=1.0\hsize}X>{\centering\arraybackslash\hsize=1.0\hsize}X>{\centering\arraybackslash\hsize=1.0\hsize}X>{\centering\arraybackslash\hsize=1.0\hsize}X>{\centering\arraybackslash\hsize=1.0\hsize}X
>{\centering\arraybackslash\hsize=1.0\hsize}X>{\centering\arraybackslash\hsize=1.0\hsize}X}
 \hline \hline  
                   & \multirow{2}{*}{SOGGA}     &  \multicolumn{4}{c}{SOGGA with percentage HF mixing}    & \multirow{2}{*}{Expt.}    \\ 
                                                               \cline{3-6}
                   &                             &  5$\%$   & 10$\%$  & 15$\%$   & 20$\%$   &      \\
\hline
$a$                &   5.565                     &  5.561   &   5.553 &    5.542 &   5.532  &  5.566   \\
                   &   (0.00)                    &  (0.08)  & (0.22)  &   (0.42) &  (0.59)  &          \\
$b$                &   5.534                     &  5.520   &   5.510 &  5.502   &  5.496   &   5.531  \\
                   &  (0.07)                     &   (0.19) & (0.37)  &  (0.52)  &   (0.62) &          \\
$c$                &   7.854                     &  7.841   &   7.827 &  7.818   &  7.820   &   7.844  \\
                   &  (0.12)                     &   (0.04) & (0.21)  &  (0.34)  &   (0.30) &          \\
$V$                &  241.90                     &  240.71  &  239.48 &  238.37  &  237.78  &   241.44 \\
                   &  (0.19)                     &  (0.30)  & (0.81)  &  (1.27)  &   (1.52) &          \\
$\phi$             &   159.97                    &  160.30  &  160.79 &  161.46  &  160.79  &   161.97 \\
                   &  (1.23)                     &   (1.03) & (0.73)  &  (0.31)  &   (0.73) &          \\
$\theta$           &   74.40                     &  75.30   &   75.90 &  76.25   &  76.80   &   77.16  \\
                   &  (3.57)                     &   (2.41) & (1.63)  &  (1.18)  &   (0.46) &          \\
Ru-O1              &  1.994                      &  1.990   &   1.985 &  1.980   &  1.983   &   1.986  \\
                   &  (0.41)                     &  (0.20)  & (0.05)  &   (0.27) &  (0.14)  &          \\
Ru-O21             &  1.998                      &  1.992   &   1.987 &  1.981   &  1.993   &   1.986  \\
                   &  (0.58)                     &  (0.28)  & (0.01)  &   (0.26) &  (0.33)  &          \\  
Ru-O22             &  1.996                      &  1.990   &   1.985 &  1.980   &  1.960   &   1.987  \\
                   &  (0.44)                     &  (0.16)  & (0.11)  &   (0.38) &  (1.37)  &          \\
O1-Ru-O21          &  90.21                      & 90.21    &  90.19  &   90.20  & 90.25    &   90.25  \\
                   &  (0.05)                     & (0.05)   & (0.06)  &    (0.06)&  (0.00)  &          \\
O1-Ru-O22          &   90.39                     &  90.46   &   90.46 &  90.45   &  90.00   &   90.31  \\
                   &  (0.08)                     & (0.17)   & (0.16)  &   (0.15) &   (0.35) &          \\
O21-Ru-O22         &   91.21                     &  91.30   &   91.28 &  91.20   &  91.05   &   91.08  \\
                   &  (0.14)                     & (0.24)   & (0.22)  &   (0.13) &   (0.04) &          \\
MARE$_1$           &  0.10                       &   0.15   &   0.41  &   0.64   &   0.76   &          \\
MARE$_2$           &  2.40                       &   1.72   &   1.18  &   0.75   &   0.59   &          \\ 
MARE$_3$           &  0.29                       &   0.18   &   0.10  &   0.21   &   0.37   &          \\ 
MARE$_\text{T}$    &  0.57                       &   0.43   &   0.38  &   0.44   &   0.54   &          \\    

\hline \hline
\end{tabularx}
\endgroup
\end{table}  

\begin{table}
\caption{\label{tab3}
Structural parameters of orthorhombic ($Pbnm$) SrRuO$_{3}$ calculated within WC framework and compared to the experimental data. Lattice constants $a$, $b$, and $c$ together with bond distances Ru-O1, Ru-O21, and Ru-O22 are given in \AA, volume $V$ is given in \AA$^{3}$, angles $\phi$, $\theta$, O1-Ru-O21, O1-Ru-O22, and O21-Ru-O22 are given in degrees. MARE (in $\%$) stands for the mean absolute relative error: MARE$_1$ is evaluated for $a$, $b$, $c$, and $V$; MARE$_2$ for $\phi$ and $\theta$; MARE$_3$ for Ru-O1, Ru-O21, Ru-O22, O1-Ru-O21, O1-Ru-O22, and O21-Ru-O22; MARE$_\text{T}$ denotes the total MARE of all 12 structural parameters. The numbers in brackets (in $\%$) represent absolute relative errors for each structural parameter.  }
\begingroup                           
\renewcommand{\arraystretch}{0.74}    
\footnotesize
\begin{tabularx}{\textwidth}{>{\hsize=1.0\hsize}X>{\centering\arraybackslash\hsize=1.0\hsize}X>{\centering\arraybackslash\hsize=1.0\hsize}X>{\centering\arraybackslash\hsize=1.0\hsize}X>{\centering\arraybackslash\hsize=1.0\hsize}X
>{\centering\arraybackslash\hsize=1.0\hsize}X>{\centering\arraybackslash\hsize=1.0\hsize}X}
 \hline \hline  
                   & \multirow{2}{*}{WC}     &  \multicolumn{4}{c}{WC with percentage HF mixing}    & \multirow{2}{*}{Expt.}    \\ 
                                                               \cline{3-6}
                   &                             &  5$\%$   & 10$\%$  & 15$\%$   & 20$\%$   &      \\
\hline
$a$                &   5.580                     &  5.573   &   5.565 &    5.556 &   5.543  &  5.566   \\
                   &   (0.26)                    &  (0.14)  & (0.01)  &   (0.18) &  (0.40)  &          \\
$b$                &   5.553                     &  5.538   &   5.526 &  5.515   &  5.531   &   5.531  \\
                   &  (0.40)                     &   (0.14) & (0.08)  &  (0.28)  &   (0.02) &          \\
$c$                &   7.877                     &  7.862   &   7.847 &  7.834   &  7.813   &   7.844  \\
                   &  (0.42)                     &   (0.23) & (0.04)  &  (0.12)  &   (0.40) &          \\
$V$                &  244.08                     &  242.69  &  241.32 &  240.05  &  239.57  &   241.44 \\
                   &  (1.09)                     &  (0.52)  & (0.05)  &  (0.58)  &   (0.78) &          \\
$\phi$             &   159.58                    &  160.03  &  160.55 &  161.15  &  159.98  &   161.97 \\
                   &  (1.47)                     &   (1.20) & (0.87)  &  (0.51)  &   (1.23) &          \\
$\theta$           &   74.21                     &  75.01   &   75.59 &  76.08   &  75.93   &   77.16  \\
                   &  (3.82)                     &   (2.79) & (2.03)  &  (1.40)  &   (1.60) &          \\
Ru-O1              &  2.001                      &  1.996   &   1.990 &  1.985   &  1.983   &   1.986  \\
                   &  (0.77)                     &  (0.51)  & (0.24)  &   (0.02) &  (0.11)  &          \\
Ru-O21             &  2.005                      &  1.999   &   1.993 &  1.987   &  1.998   &   1.986  \\
                   &  (0.95)                     &  (0.62)  & (0.32)  &   (0.04) &  (0.59)  &          \\  
Ru-O22             &  2.003                      &  1.997   &   1.991 &  1.985   &  1.976   &   1.987  \\
                   &  (0.80)                     &  (0.48)  & (0.19)  &   (0.10) &  (0.54)  &          \\
O1-Ru-O21          &  90.14                      & 90.14    &  90.14  &   90.14  & 90.42    &   90.25  \\
                   &  (0.12)                     & (0.12)   & (0.12)  &    (0.12)&  (0.19)  &          \\
O1-Ru-O22          &   90.33                     &  90.39   &   90.39 &  90.38   &  89.90   &   90.31  \\
                   &  (0.02)                     & (0.08)   & (0.09)  &   (0.08) &   (0.46) &          \\
O21-Ru-O22         &   91.21                     &  91.26   &   91.25 &  91.22   &  90.86   &   91.08  \\
                   &  (0.14)                     & (0.19)   & (0.19)  &   (0.15) &   (0.24) &          \\
MARE$_1$           &  0.55                       &   0.26   &   0.05  &   0.29   &   0.40   &          \\
MARE$_2$           &  2.65                       &   1.99   &   1.45  &   0.95   &   1.41   &          \\ 
MARE$_3$           &  0.47                       &   0.34   &   0.19  &   0.08   &   0.36   &          \\ 
MARE$_\text{T}$    &  0.86                       &   0.59   &   0.35  &   0.30   &   0.55   &          \\    

\hline \hline
\end{tabularx}
\endgroup
\end{table} 

\end{document}